\documentclass[11pt]{article}
\usepackage{mathrsfs}
\usepackage{pstricks}
\usepackage{color}
\usepackage{graphicx}
\usepackage{slashed}

\begin{document}

\begin{center}
\textbf{Formation of the seed black holes: a role of quark nuggets?}
\vskip 10mm X. Y. Lai and R. X. Xu \\ {\small {School of Physics and
State Key Laboratory of Nuclear Physics and Technology,\\ Peking
University, Beijing 100871, China\\ email: xylai4861@gmail.com}}
\end{center}

\begin{abstract}
Strange quark nuggets (SQNs) could be the relics of the cosmological
QCD phase transition, and they could very likely be the candidates
of cold quark matter if survived the cooling of the later Universe,
although the formation and evolution of these SQNs depend on the
physical state of the hot QGP (quark-gluon plasma) phase and the
state of cold quark matter. We reconsider the possibility of SQNs as
cold dark matter, and find that the formation of black holes in
primordial halos could be significantly different from the standard
scenario. In a primordial halo, the collision between gas and SQNs
could be frequent enough, and thus the viscosity acting on each SQN
would decrease its angular momentum and make it to sink into the
center of the halo, as well as heat the gas. The SQNs with baryon
numbers less than $10^{35}$ could assemble in the center of the halo
before the formation of primordial stars. A black hole could form by
merger of these SQNs, and then it could quickly become as massive as
about $10^3\ M_\odot$, if the surrounding SQNs or gas cloud is
massive enough. The black holes formed in this way could be the
seeds for the supermassive black holes at redshift as high as $z\sim
6$.

\end{abstract}


\section{Introduction}

\

QCD (quantum chromo-dynamics) is believed to describe the strong
interactions between quarks and gluons, and is characteristic by
asymptotic freedom at high energy scale and color confinement at low
energy scale.
The scale of QCD is $\Lambda_{\rm QCD}=\mathcal{O}$(100 MeV), below
which quarks and gluons are confined into hadrons.
Because the early Universe was in a hot and dense phase, in its
cooling precess, cosmological QCD phase transition took place, when
quarks and gluons confined into hadrons.
If the cosmological QCD phase transition is the first order phase
transition, there could be dramatic consequences on astrophysics.
Witten~\cite{witten84} suggested a separation of phases during the
QCD phase transition, and suggested that most of the deconfined
quarks would be concentrated in the dense and invisible strange
quark nuggets (SQNs for short, which are composed of up, down and
strange quarks).
Moreover, the ratio of mass density of SQNs to hadrons could be
approximately the ratio of mass density of dark matter to ordinary
matter.

It depends on the physical state of QGP (quark-gluon plasma) and
strange quark matter to determine the formation and evolution of
SQNs.
Unfortunately, there is no experimental result or theoretical method
which can give clear answers.
The order of cosmological QCD phase transition is still uncertain,
which could be first order, second order or crossover.
Lattice QCD calculations predict that the QCD phase transition is a
simple crossover~\cite{Aoki2009}; however, it can be first order if
the neutrino chemical potentials are sufficiently
large~\cite{Schwarz2009}.
Here we take the assumption of first order transition and study the
consequences; after all, in the other two cases, there would be no
formation of SQNs, and all of the free quarks would be confined into
hadrons.

Another important process is that in the hot universe, an SQN would
lose its baryons, depending on the state of quark matter.
This issue has been studied, in the framework of MIT bag
model~\cite{alcock85,madsen86}, the chromoelectric flux tube
model~\cite{flux}, the color superconductivity
model~\cite{horvath94} and so on.
However, the state of quark matter could be more complex, just as
showed in the experiments of relativistic heavy ion collisions
(RHIC), which indicated a strongly coupled QGP
(sQGP)~\cite{Shuryak}.
After the formation, the temperature of each SQN would decrease due
to the cooling by neutrinos, thus the interaction between quarks
could be much stronger than in the case of hot QGP.
As we will discuss in this paper, the survived SQNs would have
interesting astrophysical consequences.
If the astrophysical observations could give some constraints on the
baryon number of SQNs, we can get some hits as to the state of quark
matter at low energy scales, complementary to the terrestrial
experiments.

As a dark matter candidate, SQNs are different from the conventional
version of dark matter.
First, SQNs is a special kind of baryons with large baryon number,
so they are still in the standard model of particle physics; whereas
the conventional dark matter is of new physics.
Second, the mass scales are different.
The largest SQNs are macroscopic particles with mass $M\sim 10^{-9}
M_\odot$ and length of $\sim 1000$ cm; however, for conventional
dark matter particles, the largest mass is usually 100 GeV, and they
are always been treated as mass points.

On the other hand, although SQNs are baryon matter, they have the
characteristics of dark matter.
First of all, because of the very low electric charge per baryon
($\sim 10^{-5}$)~\cite{farhi84}, the electric charge-to-mass ratio
is very small even if they are completely ionized.
This means that the gravitational force dominates the interaction
between SQNs.
Secondly, unlike the ordinary atoms, the electrons of an SQN is in
the continuous states, thus an SQN cannot radiate via the electron
transitions  between different energy levels, which means they are
``dark''.
In addition, it is hard for an SQN to absorb protons and ordinary
atoms because of the Coulomb barrier on its surface~\cite{farhi84},
so we do not have to worry about the absorption of ordinary matter
by SQNs.
Because SQNs were non-relativistic particles at the time they
formed, they manifestly like the cold quark matter (CDM).
In this case, the overall picture of structure formation in the
standard cosmology with cold dark matter ($\Lambda$CDM) could not
change significantly, although this has not been tested by numerical
simulations.

SQNs as dark matter could not lead to conflicts with observations.
In the early Universe they would absorb neutrons and repel protons,
thus reducing primordial helium production.
Madsen~\cite{madsen85} found that if the dark matter is completely
composed of SQNs, the consistency between prediction of Big Bag
Nucleosynthesis (BBN) and observational result for the helium
abundance would not be affected as long as their radius exceed
$10^{-6}$ cm.
We revisit the issue using the more reasonable cosmology and more
accurate observational result of the helium abundance, and find that
the constraint for the radius is $R\geq 2\times 10^{-5}$ cm, a
little more severe than before.
On the other hand, SQNs in our galaxy could have been depleted by
the SNe acceleration, so it is hard for us to detect them.

SQNs as dark matter could not significantly affect the overall
scenario of structure formation in the standard $\Lambda$CDM model.
However, they may affect the formation of black holes in primordial
halos.
Because the SQNs are macroscopic particles, inside a primordial
halo, in the region where the gas density is high enough,
the SQNs could feel the viscosity due to their collisions with
gas.
This would decrease their angular momentum and make them to sink
into the center of the halo before the formation of primordial stars
(Pop III stars) in the halo, if their baryon number
$A$ satisfies $A \leq 10^{35}$.
At the same time, the gas would be heated and thus collapse more slowly.
The collision and merger of these SQNs that assemble in the center
of the halo could lead to a formation of a black hole with mass of
about $3\ M_\odot$.
The rest of SQNs will sink into the black hole through viscosity
too, and if their total mass is higher than $10^3\ M_\odot$, then a
massive black hole will quickly form.
Even if there are not enough SQNs which could sink into the center
of the halo, if the surrounding gas cloud is massive enough, the
black hole will also increase its mass to be higher than $10^3\
M_\odot$ by accreting the surrounding gas quickly, with nearly
spherical accretion rate.
Consequently, in both cases a massive black hole will form in a
short enough timescale, which could be the seed for the supermassive
black holes ($M\sim10^9 M_\odot$) at high redshifts ($z\sim 6$).
We assume that dark matter is completely composed of SQNs, but even
if they only make up a fraction of dark matter, the results could
almost be the same qualitatively.

This paper is arranged as follows.
To illustrate why SQNs could be a candidate of dark matter, we show
a brief review of their formation and evolution in Section 2.
The constrains from BBN and our present galaxy are discussed in
Section 3.
In Section 4, we show the implications for the formation of
supermassive black holes at high redshifts.
We make conclusions and discussions in Section 5.

\section{Formation and evolution of SQNs}
\

\subsection{Formation}

\

SQNs formed in the first-order QCD phase transition in the early
Universe~\cite{witten84,schwarz03}.
The baryon number inside the Hubble volume (with radius about 10 km)
is
\begin{equation}A_{\rm max}=10^{48}(\frac{170\ \rm MeV}{T_{\rm
c}})^2 ,\end{equation}
which is the maximum baryon number of SQNs, corresponding to maximum
mass of $M_{\rm max}\sim 10^{-9} M_\odot$.
Here $T_{\rm c}\simeq170$ MeV is the critical temperature of the
phase transition.
The age of the Universe is $t\sim 10^{-6}$ s.
Before the phase transition, the particles in the Universe are
quarks (u, d, and s quarks), gluons, leptons and photons.
When the temperature of the Universe reaches $T_{\rm c}$, hadron
drops nucleate by a small amount of supercooling.
The hadron bubbles grow and release latent heat to keep the
temperature of Universe at $T_{\rm c}$, and quarks outside of the
hadron bubbles are still in QGP phase.
Once hadron bubbles occupy roughly half of the total volume, they
begin to collide and merge, leaving the Universe with shrinking
bubbles of QGP immersed in a hadron matter medium.
The QGP phase could keep the temperature and pressure equilibriums
with hadron phase by the release of neutrinos, which is efficient
because the neutrinos have long mean-free path.
In this way, the QGP bubbles shrink without losing baryons (baryons
and antibaryons annihilated into neutrinos), and the baryon number
density increases.
The coexistence of the two phases can maintain until the baryon
number density of QGP phase is large enough ($\sim 3-5$ times of
nuclear matter densities), when the QGP bubbles condense into SQNs.

If we assume that at the epoch of transition, QGP phase is composed
of massless and free quarks and hadron phase is composed of
non-relativistic hadrons, we can get the baryon number density for
each phase~\cite{witten84}.
In the condition of equilibrium of the two phases, the ratio of
baryon number density in hadron phase to the QGP phase is $\epsilon
\simeq$ 0.003 to 0.27, if $T_{\rm c}=$ 100 MeV to 200 MeV.
The ratio of mass density of ordinary matter to dark matter is in
this range, which is one of the reasons to take SQNs as a candidate
of dark matter.

An SQN is composed of nearly equal numbers of up, down and strange
quarks, with a few electrons to maintain it electrically neutral.
Strange quark matter could be more stable than the ordinary
nucleus~\cite{Bodmer1971,witten84}, since the existence of strange
quarks could make the energy per baryon to be lower than that in
ordinary nucleus under reasonable QCD parameters~\cite{farhi84}.
Strange quark stars as another kind of strange quark matter in our
Universe has attracted a lot of attention since
1970's~\cite{Itoh1970}.

It is important for us to get the initial mass-spectrum of SQNs,
which depends on the properties of QGP.
Although some results were derived~\cite{india}, they were model
depended.
They treated the quark phase by MIT bag model and found that the
initial baryon numbers were between $10^{38}$ and $10^{47}$,
depending on the models for the rate of hadron nucleation.
It is conventional to apply MIT bag model to describe the quark
phase; however, recent results of Relativistic Heavy Ion Collisions
(RHIC) experiments show some evidences that the interaction between
quarks is very strong, i.e., the strongly coupled quark-gluon
plasma~\cite{Shuryak}.
This means that QGP is composed of strongly interacting quarks, and
the perturbative QCD is inadequate to describe it.
If this is the real case, QGP will be more complex than we
imagine~\cite{sQGP}, and the calculation of the initial baryon
number distribution of the quark nuggets will become more involved.
Moreover, the ratio of baryon number density of hadron phase to QGP
phase $\epsilon$ should be recalculated.

Derivation of the initial mass-spectrum of SQNs is out of the range
of our discussion in this paper, although it is important for
studying the cosmological QCD phase transition.
In addition, different initial mass-spectrum would not lead to much
differences in the final mass-spectrum after the evaporation
process.
Therefore, we assume that most of SQNs have initial baryon number
larger than $10^{38}$.

\subsection{Evolution}

\

Strange quark matter could be more stable than the ordinary nucleus.
However, in the environment of the hot early Universe SQNs suffer
the losing of baryons from their surfaces, just like liquid water
drops will evaporate.
The evaporation begins when the mean free path of the neutrinos is
larger than the size of SQNs, when heat could be transported by
neutrinos into these nuggets.
Alcock and Farhi~\cite{alcock85} found that even the largest SQNs
formed in the cosmological QCD phase transition (with baryon number
$A\simeq10^{48}$) could not survive the evaporation of hadrons in
the cooling of the Universe.
Madsen~\cite{madsen86} then pointed out that the absorption of the
nucleons dominates over evaporation when the temperature decreases
to between 10 and 20 MeV.
He also took into account the increases of the effective binding
energy on the surface of a nugget, and found that at it is possible
for SQNs with baryon number $A\geq 10^{45}$ to survive.
Further work applied chromoelectric flux tube model and considered
the effects of strong coupling constant $\alpha_{\rm
s}$~\cite{flux}, and found that a nugget having baryon number $A\geq
10^{39}$ could survive evaporation.
All of the above work treated the quark phase in the framework of
MIT bag model.
Recently, a color-superconductivity (CS) phase and a color-flavor
locked (CFL) phase have been suggested in quark matter, in which
cases the quarks pairing could form because of the weak interaction
between quarks.
Horvath~\cite{horvath94} studied the evaporation of SQNs taking into
account the QCD pairing and found that a nugget as small as $A\sim
10^{42}$ could survive evaporation.

The evaporation process depends on the state of strange quark
matter, which is difficult for us to derive theoretically or
experimentally.
The temperature inside an SQN will decrease after its formation due
to the cooling by neutrinos, so if QGP is strongly coupled at the
transition temperature, the interaction between quarks could be
stronger inside SQNs during the evaporation process.
At zero temperature, in quark nuggets with baryon density $\sim 3-5$
times of nuclear matter densities, quark clustering could occur
because of such strong interaction~\cite{xu03,xu09}.
Although the evaporation process last until the temperature of the
Universe drops at about 10 MeV, the temperature inside SQNs is much
lower than the chemical potential of quarks, which means that the
finite temperature effect is not important.
The interaction between quarks or quark-clusters could decrease the
evaporation rate, and make the nuggets easier to survive.
The detailed calculation should involve non-perturbative QCD, so it
is a difficult task.
In this paper, we try to get some constrains from astrophysics on
the baryon numbers of the survived SQNs.
Because the evaporation process of SQNs depends on the state of
quark matter, the constrains to the baryon numbers of the survived
SQNs could give us some hints as to the properties of low energy
QCD.

\section{SQNs as dark matter in the early and present Universe}

\

An SQN is composed of nearly equal numbers of u, d and s quarks,
with a few electrons to maintain it electrically neutral.
This is because the number of s quark is less than u and d quarks,
due to the larger mass of s quarks.
The binding energy of the outermost electrons is very small ($\sim$
1 eV if we describe the distribution of electrons in an SQN by the
Thomas-Fermi-Dirac model), so they were easy to be ionized in the
early Universe and in the present galaxies.

\subsection{Constrains from BBN}

\

The influences of charged dark matter on BBN have been discussed by
some authors (e.g, in~\cite{Kohri2009,Maxim2008}), but they only
considered the dark matter particles with negative charges, which
could form bound states with ordinary nucleus.
SQNs are positively charged, so they can only combine with
electrons, and they would influence BBN in a different way.

Under the MIT bag model, the electrostatic potential at the surface
of quark matter is of order 10 MeV~\cite{farhi84}.
Therefore, when the temperature of the Universe is under 10 MeV,
SQNs repel protons.
However, they absorb neutrons.
At temperature $T\gg 1$ MeV, the weak interaction is in equilibrium,
so even though SQNs are absorbing neutrons, the weak interaction can
keep the neutron-to-proton ratio at the equilibrium value.
The electroweak freeze-out take place at $T_{\rm F}\sim 1$ MeV.
From then on, the absorption of neutrons by SQNs will affect the
neutron-to-proton ratio, and consequently affect the process of BBN
which begins at the temperature $T_{\rm N}\sim 0.1$ MeV.
The standard theory of BBN has been tested very well by the
observations of CMB (Cosmic Microwave Background) and the light
element abundances.
As a candidate of dark matter, the absorption of neutrons by SQNs
should not change this consistency.
This issue was first studied by Madsen~\cite{madsen85}.

For simplicity, we suppose that all SQNs have the same radius $R$.
The absorption rate per neutron by SQNs is
\begin{equation}r=\sigma_{\rm n} v_{\rm n} n_{\rm Q},
\end{equation}
where $\sigma_{\rm n}=\pi R^2$ is the cross section for the
absorption, $v_{\rm n}=(8kT/\pi m_{\rm n})^{1/2}$ is the velocity of
neutrons of mass $m_{\rm n}$ at temperature $T$, and $n_{\rm Q}$ is
the number density of SQNs.
$n_{\rm Q}$ can be written as
\begin{equation}
n_{\rm Q}=\Omega_{\rm Q}\rho_{\rm crit,0}(\frac{T}{T_0})^3/m,
\end{equation}
where $m=10^{15}\ \rm g/cm^3$ is the mass of each SQN, $\Omega_{\rm
Q}$ is the present contribution of SQNs to the density in units of
the present critical density $\rho_{\rm crit,0}=1.88\times
10^{-29}h^2\ \rm g/cm^3$ ($100h\ \rm km s^{-1} Mpc^{-1}$ is the
present value of the Hubble parameter), and $T_0$ is the present
temperature.
Collecting all the above terms, we get
\begin{equation}r\simeq1.7\times 10^{-4}\ {\rm s^{-1}}\ (\frac{T}{1\ {\rm
MeV}})^{\frac{7}{2}}(\frac{10^{-2}{\rm cm}}{R})(\frac{\Omega_{{\rm
Q}}h^2}{0.3})(\frac{3{\rm K}}{T_0})^3a^{-1},
\end{equation}
where $a$ is a parameter of order unity giving the relation between
mass and radius of an SQN, $m=a(10^{15}{\rm g/cm^3})R^3$.

Let $n_{\rm n}$ and $n_{\rm p}$ denote the neutron and proton
abundances at the beginning of BBN, and $x^{\prime}$ and $x$ denote
the neutron-to-proton ratio at the beginning of BBN in the presence
of SQNs and the standard case respectively.
The relation between $x^{\prime}$ and $x$ is
\begin{equation}\frac{x^{\prime}}{x}=
\frac{(n_{\rm n}/n_{\rm p})_{\rm SQN}}{(n_{\rm n}/n_{\rm p})_{\rm
standard}}={\rm exp}(-\int^{t_{\rm N}}_{t_{\rm F}}r dt),
\end{equation}
where $t_{\rm N}$ and $t_{\rm F}$ are the time at the beginning of
BBN and electroweak freeze-out respectively.
Using the relation between time $t$ and temperature $T$,
\begin{equation}t\simeq (\frac{1\ {\rm MeV}}{T})^2 s,
\end{equation}
we get
\begin{equation}\int^{t_{\rm N}}_{t_{\rm F}}r dt\simeq 1.8\times
10^{-4}(\frac{T_{\rm F}}{1\ {\rm
MeV}})^{\frac{3}{2}}(\frac{10^{-2}{\rm cm}}{R})(\frac{\Omega_{{\rm
Q}}h^2}{0.3})(\frac{3{\rm K}}{T_0})^3a^{-1},
\end{equation}

The abundance of $^4\rm {He}$ is denoted by $Y_{\rm p}$, $Y_{\rm
p}=2x/(1+x)$.
Combining with the CMB observation and BBN theory, $^4\rm {He}$ is
predicted to be~\cite{cyburt03}
\begin{equation}Y_{\rm p}=0.2484^{+0.0004}_{-0.0005},
\end{equation}
while the observational result for $Y_{\rm p}$ is~\cite{fields98}
\begin{equation}Y_{\rm p}=0.238\pm0.002\pm0.005.
\end{equation}
If the presence of SQNs do not spoil the consistence of the
prediction and observation, we should let
\begin{equation}\frac{2x^{\prime}/(1+x^{\prime})}{2x/(1+x)}\geq
\frac{0.2310}{0.2488},
\end{equation}
which means that
\begin{equation}R\geq 2\times
10^{-5}\ {\rm cm}\ (\frac{T_{\rm F}}{1\ {\rm
MeV}})^{\frac{3}{2}}(\frac{\Omega_{{\rm Q}}h^2}{0.3})(\frac{3{\rm
K}}{T_0})^3a^{-1}.
\end{equation}

If dark matter is composed of SQN, then $\Omega_{{\rm Q}}h^2\simeq
0.3$, therefore the lower limit for the radius of SQNs is about
$2\times 10^{-5}$ cm.
If $R$ is larger, $x^\prime$ is closer to $x$, which means that the
influence of SQNs on standard BBN (at least for $^4\rm He$) is less
significant.
An SQN with radius of about $2\times10^{-5}$ cm has baryon number
$A\sim 10^{25}$ and mass $\sim$ 10 g.
Here we assume that all of the SQNs have the same radius for
simplicity, but it is certainly not the real case.
As long as most of the SQNs have radii larger than $2\times 10^{-5}$
cm, then the abundance of $^4\rm He$ is still in consistence with
the observations.
The productions of D, $^3\rm He$ and $^7\rm Li$ are sensitive to the
ratio of baryons to photons at the epoch of BBN $\eta_{\rm bbn}$,
and in our case that dark matter is composed of SQNs, the value of
$\eta_{\rm bbn}$ would not be different from the standard case
(without SQNs).
If the abundance of $^4\rm He$ is consistent with the observations,
then the abundances of other light elements such as D, $^3\rm He$
and $^7\rm Li$ could also be consistent with the observations, as
long as the other conditions of the Universe (such as $\eta$,
$\Omega$ and so on) are the same as in the standard case.
Although a more detailed demonstration is required, this is out of
the range of discussion in this paper~\footnote{
It is worth mentioning the presence of SQNs could lead to
inhomogeneous BBN due to the inhomogeneous distribution of
baryons~\cite{Applegate1987,Jedamzik1994}, which could provide
explanations to the inconsistent between the observed D abundance
and $^4\rm H$ and $^7\rm Li$ measurements combining with CMB
measurements}.

\subsection{SQNs in present galaxies}
\

As a candidate of dark matter, SQNs have similar properties with
charged massive particles (CHAMPs) which are essentially predicted
in physics models beyond Standard Model.
Chuzhoy \& Kolb revisited the possibility of CHAMPs as charged dark
matter in our Milky Way, and they got some constrains on the
charge-to-mass ratio based on the fact that no such particles has
been detected yet on our earth~\cite{Kolb2009}.
However, they assumed that each CHAMP has unit charge, but in our
case, each SQNs could have multi-charges.
Therefore, the critical velocity for SQNs beyond which the charged
particles can escape the galaxy by gaining more energy from SNe than
losing in Coulomb scatterings could be much smaller than that have
been derived in the case of CHAMPs.
This means that it is very easy for SQNs to escape our galaxy, and
that could be the reason why we have not detected them yet.

We suppose that at least a fraction of dark matter is composed of
SQNs, and it is also worth mentioning that the constrains on the
fraction of dark matter in CHAMPs has been derived recently by
Sanchez-Salcedo et al~\cite{Magana2010}.
They find that the fraction of the mass of galactic halo in CHAMPs
should be less than about 1\%, otherwise they cannot in pressure
equilibrium in the presence of magnetic field in the galactic disk.
Because all of the SQNs in the galaxy could have depleted due to the
very low critical velocity, then we need not to put the pressure
equilibrium condition, and the constrains on the fraction of dark
matter in SQNs could be released.

\section{The implications for supermassive black holes at high redshifts}
\

If dark matter is composed of SQNs, the overall picture of structure
formation could not be quite different from the standard
$\Lambda$CDM model.
This kind of dark matter particles were non-relativistic particles
and decoupled from the cosmological radiative background when they
formed.
Because of their extremely low electric charge-to-mass ratio, even
if they are completely ionized, the electromagnetic force acting on
them is negligible compared to the gravitational force.
Despite that the standard picture of structure formation could not
be affected significantly, there are interesting implications for
formation of supermassive black holes at high redshifts.

The supermassive black holes form by accretion and merging of seed
black holes which are the end products of the first generation of
stars (primordial stars).
Recombination occurs at $z\sim 1200$ when the ``dark ages'' of the
Universe begin.
The residual ionization of the cosmic gas keeps its temperature
locked to the CMB temperature down to a redshift of
$z\sim160$~\cite{kolb90,peebles93}.
Until this point the cosmological Jeans mass (gas plus dark matter)
is $M_{\rm J}\sim 10^5 M_\odot$, which is independent of $z$.
At lower redshifts, gas cool adiabatically and $M_{\rm J}\propto
(1+z)^{3/2}$~\cite{tan04,loeb06}.
The primordial stars form at $z\sim 20$, with mass on the order of $
10^2 M_\odot$, which reionize the Universe.
A primordial star evolves quickly and end up with a black hole which
will increase its mass by accreting the surrounding gas, but the
accretion rate is constrained by Eddington rate.
Moreover, the ionizing radiation produced by a primordial star can
photo-evaporates the surrounding gas, leaving little gas available
to be accreted by the black hole~\cite{Johnson2007}.
Therefore, even if the mass of the black hole could be of the order
100 $M_\odot$, its growth rate is limited.

The discovery of the quasar at high redshifts ($z\sim
6$)~\cite{fan03} indicates that supermassive black holes with
$M_{\rm BH}\simeq 10^9 M_\odot$ had already formed by that time.
This rapid growth of black holes is still unexplained, although
their formation could be possible under some mechanisms, e.g., the
direct collapse in pre-galactic haloes~\cite{rees06}.
The formation of primordial black holes during inflation with
spectrum covering a wide range of masses could also be possible to
lead to the supermassive black holes at high
redshifts~\cite{Maxim2005}.
Another promising way is to get more massive seed black holes, e.g.,
the direct collapse into black holes by supernova explosion under
the photodisintegration pair instability if the primordial stars are
massive enough~\cite{Heger2002}.
However, a detailed calculation of trapping of Lyman-$\alpha$
photons shows that the cooling of gas in primordial gas cloud is
efficient, which means that the formation of massive primordial
stars could be difficult~\cite{Schleicher2010}.
In the following we propose another possible mechanism that could
lead to the formation of massive seed black holes.

Dark matter virialize into gravitational bound objects through the
violent relaxation, and the density distribution could be described
by the isothermal sphere.
In the period from $z\simeq 160$ to $z\simeq 30$, baryons have
decoupled from radiation, and they collapse into structure in the
gravitational potential of the dark matter halo which formed
earlier.
Before the cooling effect become dominant, the density distribution
of gas could also be approximately described by
\begin{equation}\rho_{\rm h}(r)=\frac{v^2}{4\pi G r^2},\label{2}
\end{equation}
where the circular velocity is $v$.
The total mass of gas inside radius $r$ where the mass density
of gas is $\rho_{\rm h}$ can be written as
\begin{equation}M=\int_0^r 4\pi r^{\prime2}\rho_{\rm
h}dr^{\prime}\simeq 6\times 10^3\ M_\odot\ \left(\frac{v}{1\ {\rm
km/s}}\right)^3 \left(\frac{1\ {\rm cm^{-3}}}{n_{\rm
h}}\right)^{\frac{1}{2}},
\end{equation}
and the radius where the gas number density is 1 $\rm cm^{-3}$ is
about 30 pc.
Gas are more concentrated than dark matter particles, but the
average number density of dark matter is much more than gas, so
we could estimate the total mass of dark matter inside this radius
$r\sim30$ pc to be approximately $10^2-10^4\ M_\odot$.

SQNs are macroscopic particles, so they can collide with gas.
The time scale for an SQNs with radius $R$ and baryon number $A$ to
collide with gas with number density $n_{\rm h}$
is~\footnote{The temperature in the problem we are discussing is
less than $10^4$ K, so SQNs are neutral and we need not to consider
the Coulomb interaction of SQNs. On the other hand, the
gravitational accretion of gas onto SQNs can also be neglected:
the gravitational accretion radius of one SQN with mass $M$ and
velocity $v$ is $R_{\rm a}\approx GM/v^2$, and then the ratio of the
radius of one SQN $R$ (its baryon number is $A$) to $R_{\rm a}$ is
$R/R_{\rm a}\simeq 2\times (A/10^{42})^{-2/3}$, which is lager than
one in the situation we will consider in the following.
Consequently, we use the geometric cross section of one SQN as the
its colliding cross section with gas.}
\begin{equation}t_{\rm collide}=\frac{1}{n_{\rm h}\pi R^2
v}\simeq 10^{-3}\ {\rm s}\ \left(\frac{1\ {\rm cm^{-3}}}{n_{\rm
h}}\right)
\left(\frac{10^{35}}{A}\right)^{\frac{2}{3}}\left(\frac{1\ {\rm
km/s}}{v}\right),
\end{equation}
If $n_{\rm h}\geq 1\ \rm cm^{-3}$, the collision between these two
components could be frequent.
Under such circumstance, the velocity of each SQNs will decrease due
to viscosity.
The viscosity coefficient of gas cloud with temperature $T$ is
\begin{equation} \mu =\frac{1}{3}\rho_{\rm h}v_{\rm
s}\lambda_{\rm h}=\frac{1}{3\pi R_0^2}\sqrt{3kTm_{\rm h}},
\end{equation}
where $\lambda_{\rm h}$ is the mean free path of gas molecules, $v_{\rm
s}$ is the sound velocity, $m_{\rm h}$ and $R_0$ are mass and Bohr
radius of hydrogen ($v_{\rm s}=\sqrt{3kT/m_{\rm h}}$, $\lambda_{\rm
h}=1/\pi R_0^2 n_{\rm h}$).
We can see that $\mu$ is independent of number density of gas
$n_{\rm h}$.
The viscosity force acting on each SQN with radius $R$ and velocity
$v$ is
\begin{equation} F=-6\pi \mu v R,\label{1}
\end{equation}
which implies
\begin{equation}\frac{dJ}{dt}=-\frac{6\pi \mu R}{M} J,
\end{equation}
where $J$ is the angular momentum of one SQN.
In the primordial halo with mass of about $10^5\ M_\odot$, the
virial temperature is about 100 K~\cite{loeb06}, so we can estimate
the time scale for one SQN with baryon number $A$ to sink into the
center of the halo from initial angular momentum $J_{\rm i}$ to
finial angular momentum $J_{\rm f}$ as
\begin{eqnarray}\tau_{\rm sink}&=&\frac{M}{6\pi \mu R}\lg (J_{\rm i}/J_{\rm
f})=\frac{A R_0^2}{2R}\sqrt{\frac{m_{\rm h}}{3kT}}\lg (J_{\rm
i}/J_{\rm f})\nonumber\\&\simeq&1.6\times 10^{15}\ {\rm s}\
\left(\frac{A}{10^{35}}\right)^{\frac{2}{3}}\left(\frac{100\ {\rm
K}}{T}\right)^{\frac{1}{2}}\left(\frac{\lg (J_{\rm i}/J_{\rm
f})}{10}\right). \label{a}
\end{eqnarray}
We can see that before redshift $z\sim 30$ when the age of the universe
is about $3\times 10^{15}$ s, the angular momentum of an SQN has
decreased by over ten orders of magnitude.
On the other hand, the interaction between gas and SQNs transfers
energy from SQNs to gas, and thus gas would be heated.
The collapse of the gas would then become more slowly, although it
is difficult to estimate what temperature of the gas would
become under this heating mechanism.
As a result, the collapse of SQNs could be faster than the collapse
of gas, then there could be enough SQNs which would sink into the
center of the halo.
Because $\tau_{\rm sink}$ is independent of $\rho_{\rm h}$ as long
as the viscosity force acting on each SQNs can be described by
Eq.(\ref{1}), then even the density distribution of gas evolves
dramatically after the cooling of hydrogens becomes important, the
value of $\tau_{\rm sink}$ will not be affected~\footnote{ The
Reynolds number $\mathscr{R}$ is
\begin{equation}\mathscr{R}=\frac{2R n_{\rm h}m_{\rm
h}v}{\mu}\simeq0.2\ (\frac{A}{10^{35}})^{\frac{1}{3}}(\frac{100\
{\rm K}}{T})^{\frac{1}{2}} (\frac{n_{\rm h}}{10^{16}\ {\rm
cm^{-3}}})(\frac{v}{1\ {\rm km/s}}),
\end{equation}
so as long as SQNs are not near the central of gas cloud, the
Reynolds numbers are safely smaller than one, then the viscosity
force acting on each SQN can be written in the Stokes form as in
Eq.(\ref{1}).}.

Due to the viscosity, the orbital radius of each SQN will decrease
as the decreasing of its angular momentum.
Consequently the number density of SQNs will become higher and
higher in the center of the halo, and then it is inevitable that
they will collide with each other and merge into a larger SQN.
The total mass of SQNs which suffer the viscosity and assemble in
the center of the halo could be as high as about $10^2-10^4
M_\odot$, then it is possible that a large SQN with mass of the
order of $M_\odot$ can form, which is like a quark star.
If the mass the large SQN becomes higher than about 3 $M_\odot$,
then it will suffer gravitational instability and become a black
hole~\footnote{Although the exact value of the maximum mass of quark
stars depends on the equation of state of quark matter and is still
uncertain, it could be safely smaller than about 3
$M_\odot$~\cite{Zdunik2000,lx2009}.}.
To estimate the time for SQNs to assemble within some radius, we can
take the trajectory of each SQN to be approximately helical curves
with quasi-Keplerian motion.
Note that the angular momentum of one SQN with circular orbital
radius $r$ is $J=M\sqrt{GM_{\rm t}(r)r}$, where $M_{\rm t}$ is the
total mass within the sphere of radius $r$, so the time scale derived
in Eq.(\ref{a}) can directly transform to be a function of $r$.
Assuming that the merger of SQNs with total mass $\sim 3\ M_\odot$
is quick enough if they have assembled within radius about 30 km
(the radius of innermost stable circular orbit of $3\ M_\odot$ black
hole), we can estimate the time scale for the formation of this
black hole, if the total mass $M_{\rm t}(r)$ inside the orbital
radius $r$ of each SQNs in its sinking process will not
significantly change,
\begin{equation}\tau_{\rm bh}\simeq 5\times 10^{15}\ {\rm s}\
\left(\frac{A}{10^{35}}\right)^{\frac{2}{3}}\left(\frac{100\ {\rm
K}}{T}\right)^{\frac{1}{2}}\left(\lg \frac{r_{\rm i}/30\ {\rm
pc}}{r_{\rm f}/30\ {\rm km}}\right),
\end{equation}
where $r_{\rm i}$ and $r_{\rm f}$ are the initial and finial orbital
radius of the SQN.
The age of Universe at $z\sim 20$, when Pop III stars form in the
gas cloud in the center of the primordial halo, is about $5\times
10^{15}$, so if the baryon number $A$ of each SQN is less than
$10^{35}$, then a black hole would form before the formation of the
Pop III star.
The energy released during the collapse of $3\ M_\odot$ quark star
into a black hole would not be large enough to blew away the
surrounding gas, for the following reasons.
The strange quark matter could be more stable than ordinary nucleus,
so in the process for a quark star to collapse into a black hole,
there could be only gravitational energy release.
During this process, the rest of the quark star would be heated due
to the released gravitational energy from a growing black hole
inside it.
To get a upper limit of the released energy which can be deposited
into the surrounding gas, we take the quark star to be a black-body
with radius $R_{\rm Q}$ of about 12 km and temperature about 1 MeV
(a higher temperature would lead to effective neutrino emission
which would take away most of the energy).
The time for the quark star to become a black hole could be
estimated as $(R_{\rm Q}-R_{\rm s})/c$, where $R_{\rm s}\sim 0.9$ km
is its Schwarzschild radius, and $c$ is the velocity of light, and
then we can estimate the energy released by black-body radiation
during this process as $E_{\rm bb}=\sigma T^4 R_{\rm Q}^2 (R_{\rm
Q}-R_{\rm s})/c \sim 10^{43}$ erg, where $\sigma$ is the
Stefan-Boltzmann constant.
This energy is much lower than the gravitational binding energy of
the gas cloud in most cases.
On the other hand, because of the low electric charge-to-mass ratio
of SQNs, the impact of the radiation on motion of the surrounding
SQNs would be negligible.

The surrounding SQNs can sink into the black hole through losing
angular momentum too.
In this process, the initial orbital radius $r_{\rm i}$ is much
smaller than 30 pc, and the final orbital radius $r_{\rm f}$ is
larger than 30 km as the mass of the black hole is increasing, so
the time for one SQN to sink into the black hole should be smaller
than $10^{15}$ s.
This means that all of the surrounding SQNs can sink into the black
hole in a short enough time.
If the total mass of the surrounding SQNs is higher than $10^3\
M_\odot$, then a massive black hole with mass higher than $10^3\
M_\odot$ can form.
Moreover, the surrounding gas can also be accreted into the black
hole.
Before the formation of a Pop III star, the accretion of the gas in
protostellar cloud into the black hole can be high, since the gas is
not ionized.
The accretion of the gas into the black hole can be approximated to be
spherical accretion, and the mass accretion rate can be estimated as~\cite{yoshida06}
\begin{equation}\dot{M}=4\pi \rho_{\rm h} r^2 v_{\rm s}\sim
\frac{v_{\rm s}^3}{G},
\end{equation}
where $v_{\rm s}$ is the sound velocity of the gas cloud.
For temperature $T\sim 10^3$ K, we can get $\dot{M}\sim 10^{-3}\
M_\odot\ \rm yr^{-1}$, so the total gas with mass of about $10^3\
M_\odot$ could be accreted into the central black hole in about
$10^6$ yr, if the mass of surrounding gas cloud is about $10^3\
M_\odot$.
Therefore, the black hole with mass of about $3\ M_\odot$ that forms
before the formation of Pop III stars could quickly become a massive
black hole with mass higher than $10^3\ M_\odot$ in a short enough
time, by either eating the surround SQNs or accreting the surround
gas.
On the other hand, if the total mass of SQNs that can assemble in
the center of the halo by viscosity is much less than 100 $M_\odot$,
and in addition the surrounding gas cloud is not massive enough,
then it is difficult for the massive black to form before the
formation of Pop III stars.
In this case, the evolution of the halo could be as the same as in
the standard scenario (without SQNs).

If most of the SQNs that formed in the early Universe and then
survived the evaporation process have baryon numbers $A\leq
10^{35}$, the evolution of primordial halos could be different,
which could lead to the formation of massive black holes with mass
higher than $10^3\ M_\odot$.
These massive black holes could seed the supermassive black hole
($M\sim 10^9\ M_\odot$) at high redshifts ($z\sim 6$).
Although minimum baryon number $A$ of SQNs that can survive the
evaporation process have not been derived in reasonable models, the
picture we describe here in which the massive black holes with mass
higher than $10^3\ M_\odot$ could form quickly can be seem as a
possible explanation to the formation of the high redshift
supermassive black holes.

\section{Conclusions \& Discussions}
\

We reconsider the probability of strange quark matter being the
candidate of dark matter and suggest a possible astrophysical
consequence.
If the cosmological QCD phase transition is of first order, SQNs
could form.
Quark clustering may favor this formation.
The ratio of mass density of SQNs to hadrons could be approximately
the ratio of mass density of dark matter to the ordinary matter.
Although they have a few electrons to maintain electrically neutral,
the extremely low charge-to-mass ratio leads to the domination of
gravitational over the electromagnetic force, as well as low
radiative efficiency.
In addition, it is hard for them to radiate because their degenerate
electrons are in continuous states, unlike that in the ordinary atoms.
If they survive after the evaporation of baryons, SQNs could
provide a candidate of dark matter, and the standard BBN and the
structure formation of standard $\Lambda$CDM model could not be
affected significantly.
Certainly, future numerical simulations are necessary to study their
influences in the structure formation, which are not included in
this paper.
Although having only roughly done some estimations, we can see
that there could be interesting astrophysical consequences if SQNs
did form in the early Universe and are stable throughout the
life-time of the Universe.

One of the interesting consequences could be that these dark matter
particles could influence the formation of massive black holes in
the primordial halos, as long as the baryon number of each SQNs $A$
satisfies $A \leq 10^{35}$.
Inside a primordial halo, the collision between gas and SQNs
could be frequent enough, and thus the viscosity acting on SQNs
would decrease their angular momentum, making them to sink into the
center of the halo and heating the gas.
The collision and merger of these SQNs that assemble in the center
of the halo could lead to a formation of a black hole with mass of
about $3\ M_\odot$.
If the rest SQNs that surrounding the black hole have total mass
higher than $10^3\ M_\odot$, then a massive black hole will form
after the sinking of all of the rest SQNs, which can be very
quickly.
Even if there are not enough SQNs which could sink into the center
of the halo, in the case where the surrounding gas cloud is massive
enough, the black hole will increase its mass to be higher than
$10^3\ M_\odot$ by accreting the surrounding gas.
In both cases, before the formation of Pop III stars, a massive black
hole will form, which could be the seed for the supermassive black holes
($M\sim10^9 M_\odot$) at high redshifts ($z\sim 6$).
Although the more detailed calculations and numerical simulations
are required to get a more precise result, the rough estimates we
make here could be meaningful.
In a word, if dark matter is (partly) composed of SQNs, it could not only have
all of the properties that the conventional dark matter has, but
also help us to understand some unexplained phenomenons.
On the other hand, some more fundamental questions related to the
formation, the mass-spectrum, the evolution and the electromagnetic
properties of SQNs are still uncertain.
Is the cosmological QCD phase transition is really first order?
What is the state of matter for QGP at the critical temperature?
At lower temperature, what is the state of matter for strange quark
matter whose density is several times of the nuclear matter
densities?
Theoretically, these questions involve the non-perturbative
properties of strong interaction, thus are difficult for us now to
answer.
Experimentally, getting clear evidences that show the properties of
QGP at the critical temperature has not yet been achieved with
certainty.
It is also worth mentioning that, being a candidate of dark matter,
we can get constrains to the properties of SQNs from the
astrophysical observations, such as the influences to the CMB, the
formation of primordial stars and the masses of seed black holes.

\section*{Acknowledgments}
\

We would like to thank Prof. Zuhui Fan and Prof. Fukun Liu
(Astronomy Department) and Prof. Qingjuan Yu (KIAA) for discussions
and comments, and to acknowledge useful discussions at our pulsar
group of PKU.
We also thank an anonymous referee for valuable suggestions.
This work is supported by NSFC (10935001, 10973002) and the National
Basic Research Program of China (grant 2009CB824800).


\begin{thebibliography}{99}

\bibitem{witten84}
E. Witten, Phys. Rev. D 30 (1984) 272

\bibitem{Aoki2009}
Y. Aoki, et al JHEP 0906 (2009) 088

\bibitem{Schwarz2009}
D. J. Schwarz, M. Stuke, JCAP 11 (2009) 025.

\bibitem{alcock85}
C. Alcock \& E. Farhi, Phys. Rev. D 32 (1985) 1273

\bibitem{madsen86}
J. Medsen, et al., Phys. Rev. D 34 (1986) 2947

\bibitem{flux}
P. Bhattacharjee, et al., Phys. Rev. D 48 (1993) 4630

\bibitem{horvath94}
G. Lugones \& J. E. Horvath, Phys. Rev. D 69 (2004) 063509

\bibitem{Shuryak}
E. V. Shuryak, Prog. Part Nucl. Phys., 62 (2009) 48

\bibitem{farhi84}
E. Farhi \& R. L. Jaffe, Phys. Rev. D 30 (1984) 2379

\bibitem{madsen85}
J. Maden \& K. Riisager, Phys. Lett. 158B (1985) 208

\bibitem{schwarz03}
D. J. Schwarz, Ann. Phys. 12, No. 4 (2003) 220

\bibitem{Bodmer1971}
A. R. Bodmer, Phys. Rev. D 4 (1971) 1601

\bibitem{Itoh1970}
N. Itoh, Prog. Theor. Phys., 44 (1970) 291

\bibitem{india}
A. Bhattacharyya, et al., Phys. Rev. D 61 (2000) 083509

\bibitem{sQGP}
J. L. Nagle, Eur. Phys. J. C 49 (2007) 275

\bibitem{xu03}
R. X. Xu, Astrophys. J., 596 (2003) L59

\bibitem{xu09}
R. X. Xu, J. Phys. G: Nucl. Part. Phys., 36 (2009) 064010

\bibitem{Kohri2009}
K. Kohri \& T. Takahashi, Phys. Lett. B 682 (2009) 337

\bibitem{Maxim2008}
M. Y. Khlopov, arXiv:0806.3581 [astro-ph]

\bibitem{cyburt03}
R. H. Cyburt, Phys. Lett. B 567 (2003) 227

\bibitem{fields98}
B. D. Fields \& K. A. Olive, Astrophys. J., 506 (1998) 177

\bibitem{Applegate1987}
J. H. Applegate, C. J. Hogan \& R. J. Scherrer, Phys. Rev. D 35
(1987) 1151

\bibitem{Jedamzik1994}
K. Jedamzik, G. M. Fuller \& G. J. Mathews, Astrophys. J. 423 (1994)
50

\bibitem{Kolb2009}
L. Chuzhoy \& E. W. Kolb, JCAP 07 (2009) 014

\bibitem{Magana2010}
F. J. Sanchez-Salcedo, E. Martinez-Gomez \& J. Magana,
arXiv:1002.3145 [astro-ph.CO]

\bibitem{kolb90}
E. W. Kolb \& M. S. Turner, The Early Universe, Addison Wesley,
1990.

\bibitem{peebles93}
P. J. E. Peebles, Principles of Physical Cosmology, Princeton
University Press, 1993.

\bibitem{tan04}
J. C. Tan \& C. F. McKee, Astrophys. J., 603 (2004) 383

\bibitem{loeb06}
A. Loeb, arXiv: astro-ph/0603360.

\bibitem{Johnson2007}
J. L. Johnson \& V. Bromm, Mon. Not. R. Astron. Soc. 374 (2007) 1557

\bibitem{fan03}
X. Fan, et. al, Astron. J., 125 (2003) 1649

\bibitem{rees06}
M. C. Begelman, et al., Mon. Not. R. Astron. Soc. 370 (2006) 289

\bibitem{Maxim2005}
M. Yu. Khlopov, S. G. Rubin \& A. S. Sakharov, Astropart. Phys. 23
(2005) 265

\bibitem{Heger2002}
A. Heger \& S. E. Woosley, Astrophys. J. 567 (2002) 532

\bibitem{Schleicher2010}
D. Schleicher, M. Spaans \& S. Glover, arXiv: 1002.2850 [astro-ph]

\bibitem{Blumenthal1986}
G. R. Blumenthal, et al., Astrophys. J. 301 (1986) 27

\bibitem{Zdunik2000}
J. L. Zdunik, et al., Astron.\& Astrophys. 359, 143 (2000).

\bibitem{lx2009}
X. Y. Lai and R. X. Xu, Mon. Not. R. Astron. Soc. 398, L31 (2009).

\bibitem{yoshida06}
N. Yoshida, et al., Astrophys. J., 652 (2006) 6







\end{thebibliography}
\end{document}